\documentclass[preprint,nofootinbib]{revtex4}
\usepackage{amssymb}
\usepackage{amsmath}
\usepackage{graphicx}

\begin{document}

\title{Black holes and dark energy \\
from gravitational collapse on the brane}
\author{L\'{a}szl\'{o} \'{A}. Gergely}
\affiliation{Departments of Theoretical and Experimental Physics, University of Szeged, D%
\'{o}m t\'{e}r 9, 6720 Szeged, Hungary}

\begin{abstract}
The gravitational collapse of a pressureless fluid in general relativity
(Oppenheimer-Snyder collapse) results in a black hole. The study of the same
phenomenon in the brane-world scenario has shown that the exterior of the
collapsing \textit{dust} sphere cannot be static. We show that by allowing
for pressure, the exterior of a \textit{fluid} sphere \textit{can be static.}
The gravitational collapse on the brane proceeds according to the modified
gravitational dynamics, turning the initial nearly dust-like configuration
into a fluid with tension. The tension arises from the non-linearity of the
dynamical equations in the energy-momentum tensor, and it vanishes in the
general relativistic limit. Below the horizon the tension turns the star
into dark energy. This transition occurs right below the horizon for
astrophysical black holes and far beyond the horizon for intermediate mass
and galactic black holes. Further, both the energy density and the tension
increase towards infinite values during the collapse. The infinite tension
however could not stop the formation of the singularity.
\end{abstract}

\date{\today }
\startpage{1}
\endpage{}
\maketitle

\section{Introduction}

According \ to brane-world models our observable universe is a 4-dimensional
(4d) hypersurface (the brane) with tension $\lambda $, embedded into a
5-dimensional (5d) curved space-time (the bulk). Standard model fields act
on the brane, however gravitation spreads out into all five dimensions and
it evolves according to the 5d Einstein equation. While in the original
Randall-Sundrum (RS) second model \cite{RS2} a flat brane is embedded into
an anti de Sitter (AdS) bulk in a $Z_{2}$-symmetric way, later
generalizations have evolved into considering curved branes, embedded both
symmetrically and asymmetrically into a bulk characterized by both Weyl- and
Ricci-curvatures. A negative bulk cosmological constant has a warping
effect. The bulk can host non-standard model fields, like moduli or
dilatonic scalar fields or a radiation of quantum origin. Such models are
motivated by string /\ M-theory.

In brane-world models the projection of the 5d Einstein gravity onto the
brane leads to a 4d gravitational dynamics \cite{SMS}, which is different
from the Einstein gravity. An essential modification appears at high
energies in the form of a new source term in the effective Einstein
equation, which is quadratic in the brane energy-momentum tensor. This
source term becomes negligible at low energies. Therefore early cosmology is
modified, however late-time cosmology is not. Another modification arises
whenever the bulk has a Weyl-curvature with non-vanishing projection onto
the brane. This is known as the electric part of the bulk Weyl tensor. The
possible asymmetric embedding contributes with a third unconventional source
term in the effective Einstein equation \cite{Decomp}. So does the
non-standard model bulk matter, via the pull-back to the brane of its
energy-momentum tensor. Solar System tests, all in the weak gravity regime,
could not confirm or disprove the quadratic source term, but can put strong
limits on the other types of modifications. Quantum corrections arising from
the coupling between bulk gravity and brane matter, known as induced
gravity, were introduced originally in \cite{DGP} and presented in the most
generic covariant form (including asymmetry) in \cite{Induced}. Corrections
from higher order curvature invariants, more specifically Gauss-Bonnet type
modifications, which are motivated by the AdS/CFT correspondence, were
discussed covariantly in \cite{MT}.

Even in the simplest of these models, with the Gauss-Bonnet and induced
gravity contributions switched off, the possible brane-worlds can be of
great variety, according to the symmetries of the brane. Friedmann branes,
for example, are known to be embeddible into any of the Schwarzschild /
Reissner-Nordstr\"{o}m / Vaidya - anti de Sitter bulks (depending on the
energy-momentum in the bulk; for a systematic treatment see \cite{Decomp}).
Cosmological evolution has been extensively studied in this scenario (for
symmetric embeddings see \cite{BDEL}, \cite{MaartensLR} and references
therein; for asymmetric embeddings \cite{Decomp}, \cite{KKG} and references
therein). The matter in the bulk affects the cosmological evolution on the
brane through a "comoving mass" and a bulk pressure \cite{AT}, \cite{AT1}.

Other branes of cosmological type, like the Einstein brane \cite{Einbrane}
or the Kantowski-Sachs type homogeneous brane \cite{hombrane}, are embedded
into a vacuum bulk which is \textit{not} Schwarzschild - anti de Sitter
(SAdS).

Black hole type branes again are not embedded into a SAdS bulk. The
spherically symmetric, static black hole on the brane \cite{tidalRN} is
given by the Reissner-Nordstr\"{o}m metric of general relativity, with the
(square of the) electric charge replaced by a tidal charge. The tidal charge 
$q$ can take both positive and negative values, in contrast with the general
relativistic case, when $q=Q^{2}$ represents the square of the electric
charge, $q$ being always positive. Due to the $q$ term, the tidal charged
black hole presents $r^{-2}$ type corrections to the Schwarzschild
potential. This has to be contrasted with the $r^{-3}$ type correction to
the Schwarzschild potential \cite{RS2}, \cite{GT}-\cite{GKR}, arising in the
weak field analysis of the spherically symmetric gravitational collapse on
the brane in the original RS setup. As such corrections are related to the
electric part of the bulk Weyl curvature (the Kaluza-Klein, KK modes of
gravity), the bulk in which the tidal charged black hole can be embedded,
cannot be SAdS. Neither is the bulk containing a Schwarzschild black hole ($%
q=0$) on the brane. In fact, the bulk containing the tidal charged brane
black hole remains unknown.

Incorporating black hole singularities in brane-world models is a difficult
task. In the original, simplest RS model \cite{RS2}, containing a flat
brane, there are no black holes at all. In its curved, cosmological
generalizations no black holes can exist on the brane - except as test
particles. Among the black hole space-times of general relativity,
remarkably, the Schwarzschild solution still solves the modified
gravitational equations on the brane, under the assumptions of vacuum and no
electric~bulk Weyl source term. (This is the tidal charged brane black hole
with \thinspace $q=0$.) It was conjectured in \cite{ChRH}, that a
Schwarzschild brane can be embedded in the bulk only by extending the
singularity into the bulk. In this way one obtains a black string with
singular AdS horizon. Due to the Gregory-Laflamme instability \cite{GL} the
black string can decay into a black cigar \cite{Gregory}, although other
arguments show that under very mild assumptions, classical event horizons
cannot pinch off \cite{HorowitzMaeda}. Stable black string solutions with no
Weyl source arise in the two brane model of \cite{RS1}. Recently, the
gravity wave perturbations of such a black-string brane-world were computed 
\cite{SCMlet}. Only if the bulk contains exotic matter, the Schwarzschild
brane black hole is allowed to have regular AdS horizon \cite{KT}. In any
case the brane black hole is not localized on the brane. More generically,
in brane-worlds any event horizon on the brane can hide a singularity which
may even not be on the brane.

Rotating stationary axisymmetric black holes with tidal charge, localized on
the brane in the RS brane-world model were presented in \cite{Aliev} and
brane black hole solutions in a simple model with induced gravity were given
in \cite{Papantonopoulos}, \cite{Papantonopoulos1}. The gravitational
collapse on the brane in the presence of curvature corrections was also
studied in \cite{Papantonopoulos2}.

The formation of black holes is quite different in brane-worlds, as compared
to general relativity. This is because well-known processes from general
relativity are modified due to the unconventional brane-specific sources.
However, the "energy momentum squared" source term becomes dominant at high
energies. Such high energies are certainly occurring in the final stages of
the gravitational collapse, therefore serious modifications are to be
expected in comparison with the general relativistic gravitational collapse.

Based on the tidal charged brane black hole solution, which is set as the
exterior of the collapsing object, the gravitational collapse of a \textit{%
dust sphere} was analyzed \cite{BGM} and indeed, sharp differences as
compared to the general relativistic Oppenheimer-Snyder collapse \cite%
{OppSny} were found. First it was shown that the idealized collapse of
homogeneous KK energy density with static exterior leads to either a bounce,
a black hole or a naked singularity. This result has no counterpart in
general relativity. Second, in the \textit{absence} of the tidal charge (no
KK energy density), the vacuum surrounding the collapsing sphere of \textit{%
dust} could not be \textit{static}. This is in sharp contrast with general
relativity, where the Birkhoff theorem yields the Schwarzschild solution
outside any spherically symmetric configuration as the unique vacuum. The
non-static exterior of the collapsing brane star could be the Vaidya
radiating solution on the brane \cite{DadhichGhosh}. Moreover, this can be
regarded as an intermediate radiation layer, and matched from exterior to
the tidal charged brane black hole solution \cite{GovenderDadhich}.
According to this model, the spherically symmetric collapse on the brane is
accompanied by radiation, in contrast with general relativity.
Alternatively, in a special toy model a Hawking flux was shown to appear
from a collapsing spherically symmetric dust object on the brane \cite%
{CasadioGermani}. A related result states that the vacuum exterior of a
spherical star has radiative-type stresses, and again the exterior is not a
Schwarzschild space-time \cite{GM}. An effective Schwarzschild solution on
the brane was however shown to exist if there is energy exchange between the
bulk and the brane collapsing star \cite{Pal}.

The above mentioned results refer to collapsing spherically symmetric matter
configurations with vanishing surface pressure. But as pointed out first in 
\cite{ND}, the junction conditions on the boundary of a star in brane-world
theories do not necessarily require a vanishing pressure on the junction
surface. This is, because the multitude of source terms in the effective
Einstein equation can conspire in such a way that the \textit{effective}
pressure still vanishes on the junction surface with the vacuum exterior, in
spite of a non-vanishing ordinary pressure.

The same conclusion on the junction surface emerged in \cite{NoSwissCheese},
where the possibility of a Swiss-cheese brane-world was raised. The generic
junction conditions between the Schwarzschild vacua embedded in the FLRW
brane along spheres of constant comoving radii were exploited in \cite%
{SwissCheese} and \cite{AsymmSwissCheese} by discussing models of black
strings penetrating a cosmological brane.

In this paper we drop the assumption of vanishing surface pressure on the
boundary of the collapsing matter configuration. Obviously then the dust
model for the collapsing matter is given up. The gravitational collapse of
spherically symmetric \textit{perfect fluid} matter configuration can be
modeled by immersing a FLRW sphere into a static vacuum. Mathematically the
problem of a collapsing perfect fluid sphere immersed into a static vacuum
becomes very similar to the reverse problem of embedding of the
Schwarzschild vacua into FLRW branes.

By lifting the requirement of vanishing pressure we can achieve a static
exterior of the collapsing star even in the absence of the KK energy density
and even without inserting an intermediate radiation layer. This will be
shown in the following section. In the original RS setup the assumption of a
static exterior without tidal charge can be justified whenever the radius of
the collapsing sphere is much higher then the characteristic scale $L$ of
the extra dimension \cite{EHM}. With $L=0.1$ mm from table-top experiments 
\cite{tabletop} and for astrophysical or galactic black holes, this is
certainly the case.

In section 3 we discuss the results and comment on the formation of the
black hole and singularity. Finally, section 4 contains the concluding
remarks.

Throughout the paper we use units $G=1=c$.

\section{Spherically symmetric collapse of a fluid in a static exterior}

We choose a simple scenario, with vanishing cosmological constant on the
brane, $\Lambda =0$ (Randall-Sundrum fine-tuning). The collapsing star is
described by the FLRW metric with flat spatial sections, $k=0$. The static
vacuum exterior is the Schwarzschild metric. This simplification arises by
switching off the KK modes. Table-top experiments \cite{tabletop} on
possible deviations from Newton's law currently probe gravity at
submillimeter scales and as a result they constrain the characteristic
curvature scale of the bulk to $L=0.1$ mm. Therefore our results will apply
to collapse situations which may end in black holes with radii $r_{H}>>L$.
With this choice we focus on the effect of the non-linear source terms on
the gravitational collapse.

The junction condition of these two space-times along spheres of constant
comoving radius $\chi _{0}$ was derived in \cite{NoSwissCheese}: 
\begin{equation}
a\dot{a}^{2}=\frac{2m}{\chi _{0}^{3}}\ ,  \label{junct1}
\end{equation}%
where $a\left( \tau \right) $ is the scale factor in the FLRW metric. We
apply the above result for a stellar model with boundary surface in free
fall, given by $\chi =\chi _{0}=$constant. Then $m$ is the Schwarzschild
mass of the collapsing star. The integration of Eq. (\ref{junct1}) gives the
evolution of the scale factor of the collapsing star in comoving time $\tau $
: 
\begin{equation}
a^{3/2}=a_{0}^{3/2}-\left( \frac{9m}{2\chi _{0}^{3}}\right) ^{1/2}\tau \ .
\label{atau}
\end{equation}%
In order to describe a collapse situation we have chosen the
\textquotedblright $-$\textquotedblright\ root of Eq. (\ref{junct1}) and we
have kept the integration constant $a_{0}$, which represents the scale
factor at the beginning of the collapse (at $\tau =0$)\footnote{%
In the cosmological case the corresponding choices were \textquotedblright
+\textquotedblright\ and $a_{0}=0$, see \cite{NoSwissCheese}.}. It is
immediate to see that the collapse ends when $a=0$ is reached, after finite
time $\tau _{1}=\left( 2\chi _{0}^{3}a_{0}^{3}/9m\right) ^{1/2}$.

How is the Schwarzschild mass $m$ related to the integral of the energy
density over the volume of the star? The latter, denoted $M$, is 
\begin{equation}
M=\frac{4\pi \chi _{0}^{3}a^{3}}{3}\rho \ .  \label{mrho}
\end{equation}%
From Eq. (\ref{atau}) we easily derive 
\begin{equation}
\frac{\dot{a}^{2}}{a^{2}}=\frac{2m}{\chi _{0}^{3}\left[ a_{0}^{3/2}-\left( 
\frac{9m}{2\chi _{0}^{3}}\right) ^{1/2}\tau \right] ^{2}}=\frac{8\pi m\rho }{%
3M}\ .  \label{Hubbletau}
\end{equation}%
The stellar perfect fluid obeys the brane Friedmann equation as well: 
\begin{equation}
\frac{\dot{a}^{2}}{a^{2}}=\frac{8\pi \rho }{3}\left( 1+\frac{\rho }{2\lambda 
}\right) \ .  \label{Fried}
\end{equation}%
By comparing the two expressions for $\dot{a}^{2}$ we find the relation
between the \textquotedblright physical\textquotedblright\ mass $M$ and the
mass $m$ seen from the exterior, Schwarzschild region of the brane: 
\begin{equation}
m=M\left( 1+\frac{\rho }{2\lambda }\right) \ .  \label{mM}
\end{equation}%
Obviously, both $m$ and $M$ cannot be constants, except for the trivial case 
$m=M=0$, or in the general relativistic limit $\rho /\lambda \rightarrow 0$,
when the masses become equal. That both masses cannot be constant in the
brane-world collapsing star model was already pointed out in \cite{BGM}.

If $M$ would be constant, the energy density of the star would scale as $%
\rho \left( \tau \right) \sim a^{-3}$, cf. Eq. (\ref{mrho}). The star
therefore would consist of dust, as the pressure would vanish by virtue of
the continuity equation 
\begin{equation}
\dot{\rho}+3\frac{\dot{a}}{a}\left( \rho +p\right) =0\ .  \label{cont}
\end{equation}
For such a dust sphere the exterior cannot be static, and we arrive to the
no-go result presented in \cite{BGM}.

However, as remarked earlier, it is not compulsory to impose a vanishing
pressure in the star. The energy density of an ideal fluid \textit{with
pressure} does not evolve as $a^{-3}$. This means that $M$ varies. Then the
exterior can be held static, provided $M$ varies in the proper way.

In order to\ interpret the Schwarzschild mass $m$ in terms of the interior
metric, we transform the FLRW metric 
\begin{equation}
ds_{FLRW}^{2}=-d\tau ^{2}+a^{2}\left( \tau \right) \left[ d\chi ^{2}+\chi
^{2}\left( d\theta ^{2}+\sin ^{2}\theta d\varphi ^{2}\right) \right] \ ,
\label{FLRW}
\end{equation}%
into the standard form of spherically symmetric metrics: 
\begin{equation}
ds_{FLRW}^{2}=-e^{2\psi \left( R\right) }F\left( R\right) dT^{2}+F\left(
R\right) ^{-1}dR^{2}+R^{2}\left( d\theta ^{2}+\sin ^{2}\theta d\varphi
^{2}\right) \ .  \label{standard}
\end{equation}%
The relation between the two sets of coordinates is $T=T\left( \tau ,\chi
\right) $ and $R=R\left( \tau ,\chi \right) =a\left( \tau \right) \chi $.
Therefore $dT=\dot{T}d\tau +T^{\prime }d\chi $, and $dR=\dot{a}\chi d\tau
+ad\chi $. From the $\chi -\tau $ block of the metric we obtain: 
\begin{eqnarray}
\dot{a}^{2}\chi ^{2}+F &=&e^{2\psi }F^{2}\dot{T}^{2}\ , \\
a\dot{a}\chi &=&e^{2\psi }F^{2}T^{\prime }\dot{T}\ , \\
a^{2}\left( 1-F\right) &=&e^{2\psi }F^{2}T^{\prime 2}\ .
\end{eqnarray}%
By multiplying the first equation with the third one and eliminating the
derivatives of $T$ with the second one we obtain the metric function $F=1-%
\dot{a}^{2}\chi ^{2}$. Now we define the mass $m$ contained inside radius $%
R_{0}=a\chi _{0}$ with the metric coefficient $F$ as 
\begin{equation}
F\left( R_{0}\right) =1-\frac{2m}{R_{0}}\ ,  \label{F}
\end{equation}%
thus the mass at $\chi _{0}$ is given by 
\begin{equation}
\frac{2m}{a^{3}\chi _{0}^{3}}=\frac{\dot{a}^{2}}{a^{2}}  \label{Madot}
\end{equation}%
Finally, by applying Eq. (\ref{Fried}) the mass $m$ emerges as 
\begin{equation}
m=\frac{4\pi a^{3}\chi _{0}^{3}\rho }{3}\left( 1+\frac{\rho }{2\lambda }%
\right) \ .  \label{Mrho}
\end{equation}%
This is different from the usual relation among the mass, density and
volume, represented by Eq. (\ref{mrho}), but reduces to it in the general
relativistic limit. By applying Eq. (\ref{mrho}), we recover the relation (%
\ref{mM}) between the masses $m$ and $M$, which justifies the notation $m$
in Eq. (\ref{F}). 
\begin{figure}[tbp]
\includegraphics[height=8cm]{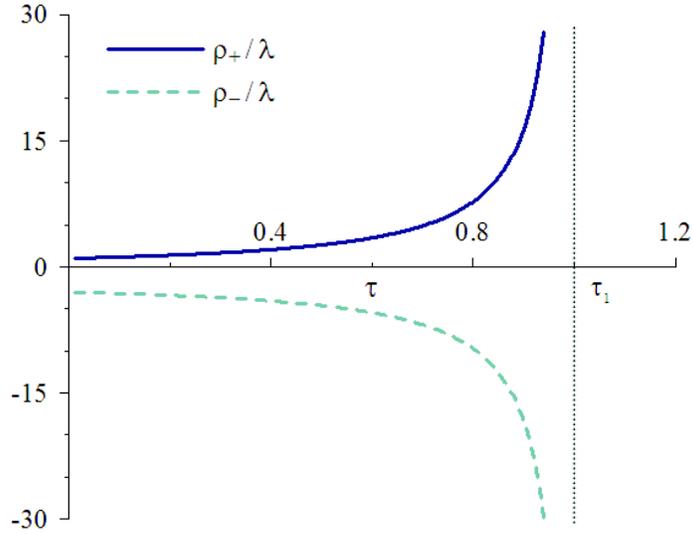}
\caption{The two branches of energy density $\protect\rho _{\pm }$ in the
collapsing star, plotted for a mass $m=2\protect\pi \protect\lambda \protect%
\chi _{0}^{3}/3$. The collapse starts at $a_{0}=1$ and the time $\protect%
\tau $ is given in units of $\left( 9m/2\protect\chi _{0}^{3}\right) ^{1/2}$
. An infinite density singularity is reached on the $\protect\rho _{+}$
branch at $\protect\tau =\protect\tau _{1}=1$.}
\label{Fig1}
\end{figure}

It is easy to show that $m$ is the quasilocal mass appearing in the
Bondi-type coordinates used by Bardeen \cite{Bardeen}. For this we need to
further transform the metric (\ref{standard}) into either the advanced or
the retarded Bardeen coordinates $\left( v,R,\theta ,\phi \right) $. The
null coordinate $v$ is given as $dv=dT+ce^{-\psi }F^{-1}dR$, with $c=\pm 1$
(the sign $+$ holds for advanced, $-$ for retarded). We obtain 
\begin{equation}
ds_{FLRW}^{2}=-e^{2\psi }Fdv^{2}+2ce^{\psi }dRdv+R^{2}\left( d\theta
^{2}+\sin ^{2}\theta d\varphi ^{2}\right) \ ,
\end{equation}%
and conclude that $m$ defined by Eq. (\ref{F}) is the Bardeen quasilocal
mass.

The mass of the star is its Bardeen quasilocal mass $m$, rather than the
\textquotedblright physical\textquotedblright\ mass $M$. It is not
surprising that these differ. The Bardeen mass contains contributions not
only from matter, but from gravitational energy as well. Therefore it should
be different compared to the general relativistic Bardeen mass (which agrees
with the \textquotedblright physical\textquotedblright\ mass $M$) because in
brane-worlds the gravitational dynamics is modified (in the present case by $%
\rho ^{2}$ source terms).

In the chosen simple scenario, with no bulk matter and no Weyl contribution
to the sources from the bulk, the effective Einstein equation for the
exterior region is the \textit{vacuum} Einstein equation of general
relativity. For spherical symmetry therefore the Birkhoff theorem applies.
Then the exterior Schwarzschild solution with mass parameter $m$ is the
unique exterior for the collapsing spherically symmetric matter
configuration. As its Schwarzschild mass $m$ agrees with the Bardeen
quasilocal mass in the star, the Bardeen mass $m$ rather than the "physical"
mass $M$ should be taken as constant.

Next, we proceed to derive $\rho \left( \tau \right) $ under the assumption $%
m$=const. This is immediate by inserting $a\left( \tau \right) $ given by
Eq. (\ref{atau}) into Eq. (\ref{mrho}). The energy density $\rho \left( \tau
\right) $ is determined by a quadratic equation 
\begin{equation}
\frac{\rho ^{2}}{\lambda ^{2}}+2\frac{\rho }{\lambda }-\frac{3m}{2\pi
\lambda \chi _{0}^{3}\left[ a_{0}^{3/2}-\left( \frac{9m}{2\chi _{0}^{3}}%
\right) ^{1/2}\tau \right] ^{2}}=0\ ,
\end{equation}%
with the solutions 
\begin{equation}
\frac{\rho _{\pm }}{\lambda }=-1\pm \sqrt{1+\frac{3m}{2\pi \lambda \chi
_{0}^{3}\left[ a_{0}^{3/2}-\left( \frac{9m}{2\chi _{0}^{3}}\right)
^{1/2}\tau \right] ^{2}}}  \label{rhopmtau}
\end{equation}%
The physical solution is $\rho _{+}$ this being positive for any $\tau <\tau
_{1}$. The energy density $\rho _{+}$ is increasing in time, reaching an
infinite value at $\tau _{1}$, at the end of collapse. During the collapse,
the \textquotedblright physical\textquotedblright\ mass $M$, scaling with $%
\left( 1+\rho /\lambda \right) ^{-1}$, decreases continuously towards zero. 
\begin{figure}[tbp]
\includegraphics[height=8cm]{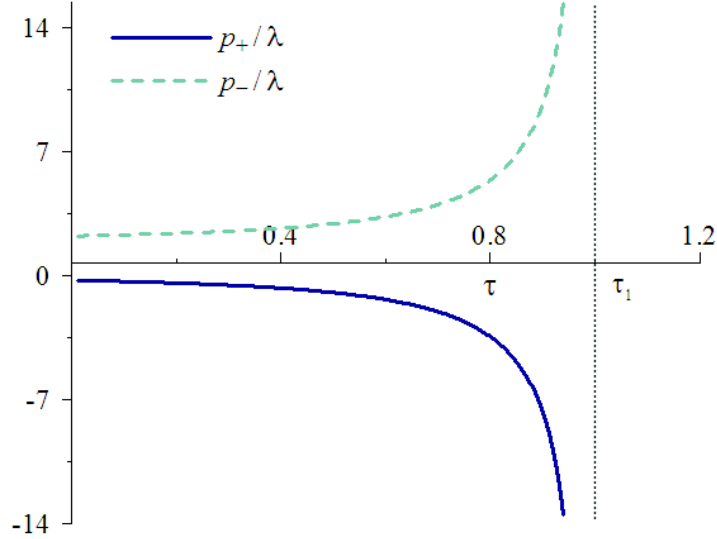}
\caption{The two branches of pressure $p_{\pm }$ in the collapsing star,
plotted as in Fig \protect\ref{Fig1}. On the physical branch $p_{+}$ is
negative during the collapse. The tension $-p_{+}$ increases to $\infty $ at 
$\protect\tau _{1}$.}
\label{Fig2}
\end{figure}

From the junction condition (\ref{junct1}) we find 
\begin{equation}
\frac{\ddot{a}}{a}=-\frac{m}{\chi _{0}^{3}a^{3}}\ ,  \label{junct2}
\end{equation}%
while the Raychaudhuri equation for the FLRW fluid sphere gives an other
expression for $\ddot{a}$: 
\begin{equation}
\frac{\ddot{a}}{a}=-\frac{4\pi }{3}\left[ \rho \left( 1+\frac{2\rho }{%
\lambda }\right) +3p\left( 1+\frac{\rho }{\lambda }\right) \right] \ .
\label{Raych}
\end{equation}%
Combining these, we obtain the equation describing the evolution of the
fluid pressure and density with radius: 
\begin{equation}
\rho \left( 1+\frac{2\rho }{\lambda }\right) +3p\left( 1+\frac{\rho }{%
\lambda }\right) =\frac{3m}{4\pi \chi _{0}^{3}a^{3}}\ .  \label{rhop}
\end{equation}%
The pressure then emerges as the algebraic root of the quadratic Eq. (\ref%
{rhop}): 
\begin{equation}
\frac{p_{\pm }}{\lambda }=1\mp \frac{1}{2}\sqrt{1+\frac{3m}{2\pi \lambda
\chi _{0}^{3}\left[ a_{0}^{3/2}-\left( \frac{9m}{2\chi _{0}^{3}}\right)
^{1/2}\tau \right] ^{2}}}\mp \frac{1}{2\sqrt{1+\frac{3m}{2\pi \lambda \chi
_{0}^{3}\left[ a_{0}^{3/2}-\left( \frac{9m}{2\chi _{0}^{3}}\right)
^{1/2}\tau \right] ^{2}}}}\ .  \label{ppmtau}
\end{equation}%
Thus our original assumption of a static exterior leads to the conclusion,
that the fluid is not dust. Time evolution of its energy density and
pressure are represented in Fig \ref{Fig1} and Fig \ref{Fig2}, respectively.

\section{Discussion}

By lifting the condition of vanishing pressure in the spherically symmetric
collapsing object we have arrived to the conclusion, that a static vacuum
exterior is possible at the price of an unexpected behaviour of the fluid.
What is the fluid composing this brane-world star consisting of? From Eqs. (%
\ref{rhopmtau} ) and (\ref{ppmtau}) we find a simpler form of the equation
describing the evolution of the pressure with density as the radius of the
star shrinks: 
\begin{equation}
\frac{p_{\pm }}{\lambda }=\frac{1}{2}\left( 1-\frac{\rho _{\pm }}{\lambda }%
\right) -\frac{1}{2}\left( 1+\frac{\rho _{\pm }}{\lambda }\right) ^{-1}\ .
\label{EOS}
\end{equation}%
This equation need not necessarily be interpreted as an equation of state
for the collapsing fluid. Rather it expresses the run of pressure with
density during the collapse. It is similar to the assumption of polytropes
as Newtonian pseudo-stellar models, where a simple power law dependence $%
p=K\rho ^{1-1/n}$ is chosen as an expression of the evolution of $p$ with
radius, in terms of the evolution of $\rho $ with radius. Such an equation
satisfies the mass equation and the equation for hydrostatic equilibrium,
but no reference to heat transfer or thermal balance are made \cite%
{StellarInteriors}. Even so, polytropes were found useful in the study of
many aspects of real stellar structure.

The relation (\ref{EOS}) was obtained by solving the brane-world
generalization for a collapse situation of the mass equation and hydrostatic
equilibrium equations, and as we will see in what follows, its
interpretation is also related to the polytropic pseudo-stellar models.

In the initial, low-energy regime of the collapse ($\left\vert \rho _{\pm
}\right\vert \ll \lambda $) the collapsing brane-world fluid approaches a
polytrope with the run of pressure versus energy density given by 
\begin{equation}
p_{\pm }\thickapprox -\frac{\rho _{\pm }^{2}}{2\lambda }\ .  \label{EOS0}
\end{equation}%
This polytrope is characterized by the constant $K=-1/2\lambda $, and
polytropic index $n=1$. The strongest bound on the minimal value of $\lambda 
$ was derived by combining the results of table-top experiments on possible
deviations from Newton's law, which probe gravity at sub-millimeter scales 
\cite{tabletop} with the known value of the 4-dimensional Planck constant.
In the 2-brane model \cite{RS1} this gives \cite{Irradiated} $\lambda
>138.59\,\,$TeV$^{4}$. A much milder limit $\lambda \gtrsim 1$ MeV$^{4}$
arises from the constraint that the dominance of the quadratic effects ends
before the Big Bang Nucleosynthesis \cite{nucleosynthesis}. From
astrophysical considerations on brane neutron stars an intermediate value $%
\lambda >5\,\times 10^{8}$ MeV$^{4}$ was derived \cite{GM}. (Note that all
these limiting values are given for $c=1=\hbar $, while the expressions of
this paper are given in units $c=1=G$. In this latter system of units the
corresponding minimal values of the brane tension are $\lambda
_{tabletop}=4.\,\allowbreak 2\times 10^{-119}$ eV$^{-2}$, $\lambda
_{BBN}=3\times 10^{-145}$ eV$^{-2}$ and $\lambda _{astro}=1.5\,\times
10^{-136}$ eV$^{-2}$, respectively.) Thus the constant becomes either a tiny
or a huge number, depending on the chosen system of units. However with
customary values of the stellar density (for the Sun $\rho _{\odot }=1408$
kg/m%
%TCIMACRO{\U{b3}}%
%BeginExpansion
${{}^3}$%
%EndExpansion
, which in units $c=1=G$ gives $\rho _{\odot }=\allowbreak 1.\,\allowbreak
8\times 10^{-150}$ eV$^{-2}$), the condition $\rho /\lambda \ll 1$ is obeyed
no matter which one of the available lower bounds and which system of units
is chosen. The collapsing fluid in the star is then indistinguishable from
ordinary dust.

At the final stage of the collapse, when $\left\vert \rho _{\pm }\right\vert
\gg \lambda $, the pressure tends to 
\begin{equation}
p_{\pm }\thickapprox -\frac{\rho _{\pm }}{2}\ .  \label{EOSfinal}
\end{equation}%
This is again a polytrope with constant $K=-1/2$ and polytropic index $%
n\rightarrow \infty $. The condition for dark energy $\rho
_{+}+3p_{+}\thickapprox -\rho _{+}/2<0$ is then satisfied on the physical
branch. This provides a mechanism of how \textit{an initial configuration of
(nearly) pressureless fluid turns into dark energy due to gravitational
collapse in the brane-world scenario}.

In the general relativistic limit ($\rho /\lambda \rightarrow 0$) the fluid
sphere degenerates into a spherically symmetric dust cloud and it remains
dust (no pressure) until the end of the collapse, which of course is the
highly idealized picture of the Oppenheimer-Snyder collapse.

The collapse is different in a brane-world. On the physical branch, as the
end of the collapse is approached ($\tau \rightarrow \tau _{1}$), the
pressure $p\rightarrow -\infty $. This means that an enormous isotropic
tension appears in the brane-world star as the collapse proceeds towards its
final stage. The role of any tension (like in solids) is to restore the
original configuration. The enormous tension appearing in the latter stages
of collapse therefore by analogy could be regarded as the backreaction of
the brane towards the stretching effect of the collapsing matter. We will
comment on the interpretation of the emerging tension later in this section.

The tension exceeding considerably the brane tension $\lambda $ is still
incapable to stop the collapse and the formation of the singularity. Why the
tension, or equivalently, the emerging dark energy is incapable to stop the
collapse, especially as $M\rightarrow 0$? The answer lies in the source
terms quadratic in the energy momentum. Towards the end of the collapse the
linear source terms in the Raychaudhuri equation (\ref{Raych}) sum up to $%
2\pi \rho _{\pm }/3$ (dark energy type source on the physical branch),
however the quadratic source terms give -$2\pi \rho _{\pm }^{2}/3\lambda $.
The latter is dominant and the collapse proceeds until the singularity is
formed.

The black hole is formed much earlier. This happens at $\tau _{H}$, when the
radius $R\left( \tau \right) =a\left( \tau \right) \chi _{0}\ $of the
collapsing fluid sphere reaches the horizon, which is at $r_{H}=2m$. From
Eq. (\ref{atau}) we get%
\begin{equation}
\tau _{H}=\frac{4m}{3}\left[ \left( \frac{a_{0}\chi _{0}}{2m}\right) ^{3/2}-1%
\right] ~.
\end{equation}%
The energy density and pressure of the fluid at horizon crossing is, cf.
Eqs. (\ref{rhopmtau}) and (\ref{ppmtau}):%
\begin{eqnarray}
\frac{\left( \rho _{\pm }\right) _{H}}{\lambda } &=&-1\pm \sqrt{1+\frac{3}{%
16\pi \lambda m^{2}}}~, \\
\frac{\left( p_{\pm }\right) _{H}}{\lambda } &=&1\mp \frac{1}{2}\sqrt{1+%
\frac{3}{16\pi \lambda m^{2}}}\mp \frac{1}{2\sqrt{1+\frac{3}{16\pi \lambda
m^{2}}}}~.
\end{eqnarray}%
Therefore%
\begin{equation}
\frac{\left( \rho _{\pm }\right) _{H}+3\left( p_{\pm }\right) _{H}}{\lambda }%
=2\mp \frac{1}{2}\sqrt{1+\frac{3}{16\pi \lambda m^{2}}}\mp \frac{3}{2\sqrt{1+%
\frac{3}{16\pi \lambda m^{2}}}}~.  \label{dark_horizon}
\end{equation}%
This expression, on the physical branch is positive between the roots $%
\left( \lambda m^{2}\right) _{1}\rightarrow \infty $ and $\left( \lambda
m^{2}\right) _{2}=3/128\pi $ and negative for any $\lambda m^{2}<3/128\pi $.

For astrophysical or galactic black holes the dark energy condition could be
obeyed only below the horizon. From Eq. (\ref{rhop}) we can find the
density-radius relation, at which $\rho +3p=0$ occurs: 
\begin{equation}
\frac{\rho ^{2}}{\lambda }=\frac{3m}{4\pi r_{de}^{3}}\ .
\label{density-radius}
\end{equation}%
Combined with Eqs. (\ref{atau}) and (\ref{rhopmtau}), this gives the radius
of dark energy crossing%
\begin{equation}
r_{de}=\frac{A^{1/3}}{\mu ^{2/3}}r_{H}~,
\end{equation}%
with $\mu $ the mass of the black hole expressed in units of solar masses $%
M_{\odot }$ $=1.\,\allowbreak 115\,4\times 10^{66}$ eV (for$\,c=1=G$) and%
\begin{equation}
A=\frac{3}{128\pi \lambda M_{\odot }^{2}}  \label{A}
\end{equation}%
a dimensionless constant.

With the astrophysical limit set on the brane tension, which in units $%
\,c=1=G$ is $\lambda _{astro}=1.5\,\times 10^{-136}$ eV$^{-2}$ we obtain $%
A=40$ (from $\lambda M_{\odot }^{2}=1.\,\allowbreak 866\,2\times 10^{-4}$)
and 
\begin{equation}
r_{de}=3.\,\allowbreak 42\mu ^{-2/3}r_{H}=\mu ^{1/3}\times \allowbreak
7.\,\allowbreak 629\,3\times 10^{66} \text{eV}.  \label{rde}
\end{equation}%
The first expression shows that the bigger the black hole, the more has the
fluid to collapse below the horizon before the dark energy condition is
obeyed. For example, for astrophysical black holes with $\mu =10,~100~$and
galactic black holes with $10^{4},10^{6},~10^{8}$ the ratio $r_{de}/r_{H}$
is $\allowbreak 0.737,~\allowbreak 0.159$ and $7.37\times
10^{-3},~3.42\times 10^{-4},~\allowbreak 1.59\times 10^{-5}$ respectively.

The second expression in Eq. (\ref{rde}) shows that the radius where the
dark energy condition is obeyed increases with the cubic root of $\mu $. For
the above examples $r_{de}$ expressed in units $10^{67}$ eV takes the values 
$1.64,~\allowbreak 3.\,\allowbreak 54$ and $16.44,~\allowbreak
76.29,~\allowbreak 354.1\,2$.

We note from the Eqs. (\ref{density-radius})-(\ref{A}) that the transition occurs
at the extreme high density $\rho =2\lambda $, where the perfect fluid
approximation may break down. \ For comparison with the known lower limits
for the brane tension $\lambda _{tabletop}=4.\,\allowbreak 2\times 10^{-119}$
eV$^{-2}$, $\lambda _{BBN}=3\times 10^{-145}$ eV$^{-2}$ and $\lambda
_{astro}=1.5\,\times 10^{-136}$ eV$^{-2}$, we give here the density range
for the densest known stellar objects, the neutron stars, extending from $%
\rho _{ns}=8\times 10^{16}$ up to $2\times 10^{18}$ kg/m%
%TCIMACRO{\U{b3}}%
%BeginExpansion
${{}^3}$%
%EndExpansion
, which in units $c=1=G$ give $\rho _{ns}^{\min }=\allowbreak 1\times
10^{-136}$ eV$^{-2}$ and $\rho _{ns}^{\max }=\allowbreak 2.\,\allowbreak
6\times 10^{-135}$ eV$^{-2}$. We see that the astrophysical limit is of the
same order of magnitude than the density of the neutron stars, which in turn
is comparable to the density of the atomic nucleus.

What would be the mass of a black hole for which the dark energy crossing
occurs exactly on the horizon? From Eq. (\ref{dark_horizon}) this is the
root $m=\left( 3/128\pi \lambda \right) ^{1/2}=A^{1/2}M_{\odot }=\allowbreak
6.\,\allowbreak 32~M_{\odot }$. The same result stems out from Eq. (\ref{rde}%
) by imposing $r_{de}/r_{H}=1$. Remarkably, this is about the minimal mass
required for black hole formation, right above the maximally allowed mass
for neutron stars, given by the Tolman-Oppenheimer-Volkoff limit $%
m_{ns}^{\max }=1.5\div 3~M_{\odot }~$\cite{TOV1}-\cite{TOV3}.

Whenever $\lambda m^{2}\gg 1$ can be assumed (thus $\mu \gg 73$), to second order in the small
parameter $\left( \lambda m^{2}\right) ^{-1}$ on the physical branch 
\begin{eqnarray}
\rho _{H} &=&\frac{3}{32\pi m^{2}}\left( 1-\frac{3}{64\pi \lambda m^{2}}%
\right) ~, \\
p_{H} &=&-\frac{9}{2048\pi ^{2}\lambda m^{4}}~,
\end{eqnarray}%
and%
\begin{equation}
\rho _{H}+3p_{H}=\frac{3}{32\pi m^{2}}\left( 1-\frac{3}{16\pi \lambda m^{2}}%
\right) \text{~.}
\end{equation}%
This barely differs from the general relativistic value, as the second,
brane-world induced term represents a tiny negative correction. The
composition of the collapsing fluid remains basically dust at horizon
crossing. For galactic black holes thus $\rho +3p>0$ everywhere above the
horizon and also in the greatest part of the collapse below the horizon.

It is to be expected that due to the existence of a fundamental length
scale, the Planck length (which in units $c=1=G$ is $L_{P}=\allowbreak
1.\,\allowbreak 221\times 10^{28}\,$eV), the brane model as presented here,
with the brane as an infinitesimal hypersurface, will break down. Rather, a
finite thickness should be assumed for the brane. Such thick brane models
were already studied, either with a scalar field on the thick brane \cite%
{scalar1}, \cite{scalar2} or an other matter form with transverse pressure
component \cite{transverse1}, \cite{transverse2}, meant to replace the brane
tension. We expect that the emergence and increase without bounds of the
tension in the collapsing fluid (\ref{EOS}) could be derived in a suitable
limit of some thick brane model, from the interaction of the fluid with the
matter configuration of the thick brane.

\section{Concluding Remarks}

We have considered a simple brane-world model containing a collapsing,
spherically symmetric stellar configuration on the brane. By considering
only objects with gravitational radius much higher than the characteristic
curvature scale of the bulk we have supressed the KK modes. Then the
exterior of the collapsing star is described by the general relativistic
vacuum Einstein equation and due to the Birkhoff theorem the exterior should
be the Schwarzschild solution.

Under these circumstances we have shown that the gravitational collapse of a
near dust-like perfect fluid star can occur with a static exterior. Due to
the modified brane-world dynamics a tension rises during the collapse, and
the "physical" mass of the star gradually diminishes, while its
Schwarzschild mass, which coincides with the Bardeen quasilocal mass, stays
constant.

For large collapsing objects, leading to galactic or intermediate mass black
holes, the collapsing fluid even at horizon crossing can be considered to a
remarkably good accuracy as dust. Tensions rise slowly, so that below the
horizon the fluid remains ordinary matter for a long time, but in the very
latest stages of the collapse it turns into dark energy.

By contrast, for the lowest mass astrophysical black holes (with the
astrophysical limit set for the brane tension), an important tension appears
already at the horizon crossing so that the dark energy condition could be
obeyed shortly after the horizon crossing.

This peculiar behaviour of the collapsing star stays hidden below the
horizon, such that a distant outside observer will sense nothing but an
usual black hole with mass $m$. In spite of the dark energy condition being
satisfied somewhere below the horizon for all black holes, the collapse of
the fluid will not be stopped or even slowed down. Due to the energy-squared
source term, appearing in the modified brane dynamics, the fluid further
evolves into a central singularity. Thus as in general relativity, the
singularity is formed in brane-world theory too.

The appearance of the tension in the collapsing fluid is a pure brane-world
effect. As we have disregarded the Weyl-source term of the bulk, the rising
tension is due solely to the source term quadratic in the fluid
energy-momentum tensor. This source term modifies the early cosmology
(before BBN) in an essential way. The same happens in the case of the
gravitational collapse: a tension rises, induced by the non-linearity of the
dynamical equations in the energy-momentum tensor. In this sense, the
tension in the fluid is an expressions of the interaction of the fluid with
the brane.

However as long as the brane is considered as a hypersurface in the
mathematical sense (with only one characteristics, the brane tension) such
an interaction could not be described phenomenologically. This feature of
the model may be an indication for the need of a description based on
thick-branes.

\begin{acknowledgments}
This work was supported by OTKA grants no. T046939, TS044665, and the J\'{a}%
nos Bolyai Fellowship of the Hungarian Academy of Sciences.
\end{acknowledgments}

\end{document}